\documentclass[11pt,leqno]{article}

\usepackage[english]{babel}

\usepackage[letterpaper,top=2cm,bottom=2cm,left=2.6cm,right=2.6cm,marginparwidth=1.75cm]{geometry}

\usepackage[hyphenbreaks]{breakurl}
\usepackage{amsmath}
\usepackage{xspace} 
\usepackage{graphicx}
\usepackage[numbers]{natbib}
\usepackage[colorlinks=true, allcolors=blue]{hyperref}
\usepackage[usenames,dvipsnames]{xcolor}
\usepackage[colorinlistoftodos]{todonotes}
\usepackage{verbatim}

\def\PROJECT{\textbf{PEACE} Project\xspace}

\title{The Case for Anticipating Undesirable Consequences of Computing Innovations Early, Often, and Across Computer Science}
\author{Rock Yuren Pang, Dan Grossman, Tadayoshi Kohno, Katharina Reinecke \\
\small{Paul G. Allen School of Computer Science and Engineering, University of Washington Seattle, WA, USA}}
\date{}

\begin{document}
\maketitle

%%%%%%%%%%%%%%%%%%%%%%%%%%%%%%%%%%%%%%%%%%%%%%%%%%%%%%%%%%%%%%%%%%%%%%%%%%%%%%%%%%%%
%%%%%%%%%%%%%%%%%%%%%%%%%%%%%%%%%%%% Paragraph 1 %%%%%%%%%%%%%%%%%%%%%%%%%%%%%%%%%%%
%%%%%%%%%%%%%%%%%%%%%%%%%%%%%%%%%%%%%%%%%%%%%%%%%%%%%%%%%%%%%%%%%%%%%%%%%%%%%%%%%%%%
From smart sensors that infringe on our privacy to neural nets that portray realistic imposter deepfakes, our society increasingly bears the burden of negative, if unintended,  consequences of computing innovations.
As the experts in the technology we create, Computer Science (CS) researchers must do better at anticipating and addressing these undesirable consequences proactively. Our prior work showed that many of us recognize the value of thinking preemptively about the perils our research can pose~\cite{do2023important}, yet we tend to address them only in hindsight. How can we change the culture in which considering undesirable consequences of digital technology is deemed as important, but is not commonly done?

%%%%%%%%%%%%%%%%%%%%%%%%%%%%%%%%%%%%%%%%%%%%%%%%%%%%%%%%%%%%%%%%%%%%%%%%%%%%%%%%%%%%
%%%%%%%%%%%%%%%%%%%%%%%%%%%%%%% Paragraph 2 %%%%%%%%%%%%%%%%%%%%%%%%%%%%%%%%%%%%%%%%
%%%%%%%%%%%%%%%%%%%%%%%%%% Barriers found in prior work %%%%%%%%%%%%%%%%%%%%%%%%%%%%
%%%%%%%%%%%%%%%%%%%%%%%%%%%%%%%%%%%%%%%%%%%%%%%%%%%%%%%%%%%%%%%%%%%%%%%%%%%%%%%%%%%%
%CS researchers often want to consider potential unintended consequences---it's just incredibly hard. 
There are many factors keeping researchers from thinking about potential social impacts in advance: the usual (and often true) lack of time argument as our move-fast mentality and deadlines take precedence over all else; the it's-someone-else's responsibility position as we look to others we consider more capable of managing this issue (e.g., the IRB, other fields, those commercializing our ideas,  other team members); the lack of formal processes and guidelines to think through potential negative effects; and the difficulties accessing diverse perspectives that could shed light on what disparate or undesirable impacts may be~\cite{do2023important}. 
When CS researchers do think about potential undesirable consequences, it is usually in hindsight, e.g., after a publication venue requires an ethics statement or after a research innovation raises concerns. Addressing consequences at such times is, of course, too late to change designs or pivot on proposed features. Sometimes it may even mean that the damage cannot be undone. As Pillai et al.\ suggest, “[u]nless ethics is integrated into every aspect of the design process and educational curriculum, it is bound to be an afterthought and thus inadequate in identifying and addressing ethical issues.”~\cite[p.2]{g2021co}.

%%%%%%%%%%%%%%%%%%%%%%%%%%%%%%%%%%%%%%%%%%%%%%%%%%%%%%%%%%%%%%%%%%%%%%%%%%%%%%%%%%%%
%%%%%%%%%%%%%%%%%%%%%%%%%%%%%%%%%%%%% Paragraph 3 %%%%%%%%%%%%%%%%%%%%%%%%%%%%%%%%%%
%%%%%%%%%%%%%%%%%%%%%%%%%%%%%%%%%%% Past Approaches %%%%%%%%%%%%%%%%%%%%%%%%%%%%%%%%
%%%%%%%%%%%%%%%%%%%%%%%%%%%%%%%%%%%%%%%%%%%%%%%%%%%%%%%%%%%%%%%%%%%%%%%%%%%%%%%%%%%%
To change this, some CS researchers have proposed and tried different approaches over the past decades. They include developing tools and methods for examining societal impacts~\cite{friedman1996value, baumer2018would} and, more recently, requiring researchers to submit an ethics statement with paper submissions~\cite{NeurIps2021} or undergo an Ethics and Society Review before grant proposals are approved and funded~\cite{berstein2021esr}. Such efforts likely raise our awareness about this issue. However, our research findings show that these actions remain insufficient to make meaningful headway on an intractable problem that computing researchers rarely anticipate and address undesirable consequences systematically in advance.

%%%%%%%%%%%%%%%%%%%%%%%%%%%%%%%%%%%%%%%%%%%%%%%%%%%%%%%%%%%%%%%%%%%%%%%%%%%%%%%%%%%%
%%%%%%%%%%%%%%%%%%%%%%%%%%%%%%%%%%%%%% Paragraph 4 %%%%%%%%%%%%%%%%%%%%%%%%%%%%%%%%%
%%%%%%%%%%%%%%%%%%%%%%%%%%%%%%%%%%% Our  Proposals %%%%%%%%%%%%%%%%%%%%%%%%%%%%%%%%%
%%%%%%%%%%%%%%%%%%%%%%%%%%%%%%%%%%%%%%%%%%%%%%%%%%%%%%%%%%%%%%%%%%%%%%%%%%%%%%%%%%%%
\vspace{\baselineskip}
\noindent After much research into and thinking about this topic, here is what we propose for moving forward:

\begin{itemize}

%%%%%%%%%%%%%%%%%%%%%%%%%%%%%%%%%%%%%%%%%%%%%%%%%%%%%%%%%%%%%%%%%%%%%%%%%%%%%%%%%%%%
%%%%%%%%%%%%%%%%%%%%%%%%%%%%%%%%%%%%% Item 1 %%%%%%%%%%%%%%%%%%%%%%%%%%%%%%%%%%%%%%%
%%%%%%%%%%%%%%%%%%%%%%%%%%%%%%%%%%%%%%%%%%%%%%%%%%%%%%%%%%%%%%%%%%%%%%%%%%%%%%%%%%%%
\item \textbf{Support research subfields across CS:}
Recent efforts to introduce ethical considerations in the technology development process have most often focused on artificial intelligence (AI) instead of encouraging all computing researchers to consider potential societal effects. This is despite the fact that many CS subcommunities have seen their own share of sometimes severe unintended consequences (e.g., in virtual reality~\cite{tseng2022dark}, sensing~\cite{Guo_2023}, haptics~\cite{cornelio2023repsonsible}, social networks~\cite{kramer2014experimental}, and accessibility~\cite{jang2014a11y}). Pointing our fingers at AI can risk that researchers in other fields feel ethical and societal impact considerations are ``someone else's problem.''~\cite{do2023important} It may also cause us to overlook opportunities for holistic improvements across the broader CS field. Instead, we believe that all computing researchers, no matter their subdiscipline, should be supported in learning about what ethics in computing means as well as how to proactively consider undesirable consequences in our work.

%%%%%%%%%%%%%%%%%%%%%%%%%%%%%%%%%%%%%%%%%%%%%%%%%%%%%%%%%%%%%%%%%%%%%%%%%%%%%%%%%%%%
%%%%%%%%%%%%%%%%%%%%%%%%%%%%%%%%%%%%% Item 2 %%%%%%%%%%%%%%%%%%%%%%%%%%%%%%%%%%%%%%%
%%%%%%%%%%%%%%%%%%%%%%%%%%%%%%%%%%%%%%%%%%%%%%%%%%%%%%%%%%%%%%%%%%%%%%%%%%%%%%%%%%%%
\item \textbf{Encourage early considerations:} Considering undesirable consequences should start when formulating the research problem to increase the likelihood that it is still possible to pivot.  
Waiting until the submission or publication stage to reflect on potential social impacts is too late to substantially address the problem and make changes to a research project. This is both because people become invested in an idea once they have put in some work, and because modifying existing innovations is considerably more time-consuming than doing so early on.

%%%%%%%%%%%%%%%%%%%%%%%%%%%%%%%%%%%%%%%%%%%%%%%%%%%%%%%%%%%%%%%%%%%%%%%%%%%%%%%%%%%%
%%%%%%%%%%%%%%%%%%%%%%%%%%%%%%%%%%%%% Item 3 %%%%%%%%%%%%%%%%%%%%%%%%%%%%%%%%%%%%%%%
%%%%%%%%%%%%%%%%%%%%%%%%%%%%%%%%%%%%%%%%%%%%%%%%%%%%%%%%%%%%%%%%%%%%%%%%%%%%%%%%%%%%
\item \textbf{Encourage regular considerations:} Just as research itself is a constantly evolving endeavor, considering undesirable consequences can also introduce unforeseen challenges that require ongoing feedback and re-evaluation. Rather than making it a one-time exercise, researchers should routinely think about the potential societal implications of their work. 
Achieving this will require changing the institutional culture and support system such that researchers are incentivized to regularly think about societal implications and learn how to do this efficiently.

%%%%%%%%%%%%%%%%%%%%%%%%%%%%%%%%%%%%%%%%%%%%%%%%%%%%%%%%%%%%%%%%%%%%%%%%%%%%%%%%%%%%
%%%%%%%%%%%%%%%%%%%%%%%%%%%%%%%%%%%%% Item 4 %%%%%%%%%%%%%%%%%%%%%%%%%%%%%%%%%%%%%%%
%%%%%%%%%%%%%%%%%%%%%%%%%%%%%%%%%%%%%%%%%%%%%%%%%%%%%%%%%%%%%%%%%%%%%%%%%%%%%%%%%%%%
\item \textbf{Support CS researchers at all levels:} Many prior approaches place most of the responsibility on a single person, namely, the person submitting a paper or the PI submitting a proposal. This can lead other research team members, such as undergraduates, graduates, postdoctoral researchers, or other collaborators, to overly rely on this one person. In fact, our prior work suggests that some faculty rely on the experience of their ``more ethics-educated'' students, while students may rely on the experience of their senior researchers and faculty, pointing to their experience~\cite{do2023important}. To break this cycle of deferred responsibility, every one of us on a research team should play a role and be supported in addressing this issue.

\end{itemize}

%%%%%%%%%%%%%%%%%%%%%%%%%%%%%%%%%%%%%%%%%%%%%%%%%%%%%%%%%%%%%%%%%%%%%%%%%%%%%%%%%%%%
%%%%%%%%%%%%%%%%%%%%%%%%%%%%%%%%%%%%% Summary %%%%%%%%%%%%%%%%%%%%%%%%%%%%%%%%%%%%%%
%%%%%%%%%%%%%%%%%%%%%%%%%%%%%%%%%%%%%%%%%%%%%%%%%%%%%%%%%%%%%%%%%%%%%%%%%%%%%%%%%%%%
In summary, we believe that any computing researcher---no matter whether their work centers around AI, human-computer interaction,  theory, or another CS subdiscipline and no matter what role they play in a research team---should be supported in considering potential undesirable consequences of their research innovations early and often in the project development cycle. This will front-load considerations of societal implications in the research process and empower people to grow their moral muscle over time. 
As creators \textit{and} potential users of our innovations, we owe it to ourselves and to each other to not repeat mistakes of the past as we transform our future.

%%%%%%%%%%%%%%%%%%%%%%%%%%%%%%%%%%%%%%%%%%%%%%%%%%%%%%%%%%%%%%%%%%%%%%%%%%%%%%%%%%%%
%%%%%%%%%%%%%%%%%%%%%%%%%%%%%%%%%%%%%% Second Half %%%%%%%%%%%%%%%%%%%%%%%%%%%%%%%%%
%%%%%%%%%%%%%%%%%%%%%%%%%%%%%%%%%%%%%%%%% UW CSE %%%%%%%%%%%%%%%%%%%%%%%%%%%%%%%%%%%
%%%%%%%%%%%%%%%%%%%%%%%%%%%%%%%%%%%%%%%%%%%%%%%%%%%%%%%%%%%%%%%%%%%%%%%%%%%%%%%%%%%%
\subsection*{What we are doing to support this vision}

At the Paul G. Allen School of Computer Science \& Engineering at the University of Washington, we are trying to turn these ideas into action. Through the \PROJECT (Proactively Exploring and Addressing Consequences and Ethics), funded by the National Science Foundation under grant \#2315937, we aim to transform our institutional culture so that any researcher gets support to routinely and proactively address undesirable consequences of their research innovations. As one of the largest top CS programs in the country, we hope the \PROJECT will transform the practices of more than 90 faculty members and their labs, their courses, postdocs, researchers, students, and collaborators at UW and in the tech industry. 

Importantly, our goal is not to inhibit emerging and risky research, but to put it in a safer context. Indeed, such research could, if done thoughtfully and cautiously, lead to a strong positive impact. Akin to the classic debate over dual-use technologies, many research innovations have both positive and negative effects on society (social media, deep fakes, and accessibility are notable examples). Identifying the negative effects can help us invest in further research and preventative mechanisms, from issuing disclaimers and regulations to reformulating research questions and directions. 

Our approach includes support for learning about undesirable consequences, anticipating them for one's own projects, and seeking advice. For example, with \textbf{case studies of past societal impacts} of computing research, we aim to highlight previous undesirable consequences in all CS subfields and how they were addressed. With \textbf{resources that scaffold brainstorming about societal impacts}, we aim to lower common barriers, such as not having the time or energy to spend on this. By providing \textbf{access to a diverse ethics board}, we further increase the support for those who may have additional questions. 

At the heart of our approach is a \textbf{PEACE Report} that lets researchers collect and share potential undesirable consequences and ways to address them as a project evolves. Inspired by the idea behind preregistering a  research study, the goal of our PEACE Report is to provide scaffolding for thinking through a project's societal impacts. If desired, it can be a shareable artifact that researchers may attach to papers, grant applications, or annual review reports. Inviting other researchers in the Allen School to review their notes may also increase awareness of other people's work and foster collaboration across research groups. 

These proposed measures would likely not succeed without some incentives and strong encouragement. Our approach is more carrot than stick: We plan to encourage students and faculty to integrate their PEACE Report into our annual review processes and to include the ethics resources in seminars for new Ph.D. students and undergraduate researchers, as well as in regular learning sessions for faculty. Finally, we will experiment with publicity and reward incentives, such as prizes for the most comprehensive and reflective report. 

While we are optimistic about the potential of our approach, we do not yet know whether this approach will adequately support researchers in considering undesirable consequences. Likewise, we have yet to investigate to what extent our approach will be adopted among researchers. It is also unclear whether the success of the project in the Allen School means it will also succeed in other institutions. 
Over the next few years, we will iteratively develop and evaluate our approach and report the results to the CS community. We warmly welcome ideas, feedback, and collaboration.

\subsubsection*{Acknowledgments}
This work was partially supported by NSF grants \#2315937 and \#2006104. We thank Kristin Osborne, Sandy Kaplan, and Magda Balazinska for their feedback on earlier versions of this article. We are also thankful for past and ongoing discussions with many people in the research community, which have significantly shaped our views. Finally, we are grateful to Timothy Brown, Ryan Calo, Priti Ramamurthy, Daniela Roesner, and Adam Romero for agreeing to support this work as part of the ethics advisory board. 
\clearpage

\bibliographystyle{ACM-Reference-Format}
\bibliography{bibliography}

\end{document}